# Impulsive Sound Detection by a Novel Energy Formula and its Usage for Gunshot Recognition

Yüksel Arslan
Elektrik ve Elektronik Mühendisliği Bölümü
Yıldırım Beyazıt Üniversitesi
Ankara, Türkiye
d1000004149@ybu.edu.tr

*Abstract*— There are many methods proposed for the detection of impulsive sounds in literature. Most of them are complex and require adaptation to ambient noise. In this paper we propose a very simple and efficient method to detect impulsive sounds. Although we use energy like most of the others to determine impulsive sounds, the way we calculate the energy is quite different. Also our calculation is immune to ambient noise and does not require any limit or adaptation. We could detect impulsive sounds embedded in various kinds of noises by using this formula.

As our ultimate aim is to detect gunshots, next phase of impulsive sound detection is gunshot recognition phase. Detected impulsive sounds are fed into recognition phase in which we can decide on gunshots with high success rate.

## I. INTRODUCTION

The impulsive sound detection algorithm described here can be used generally but our concern is to detect gunshots. (In literature recognition and classification are used interchangeably, we prefer recognition.) A gunshot recognition system has two phases (Figure-1):

  a) *Impulsive sound detection*
  b) *Gunshot recognition*

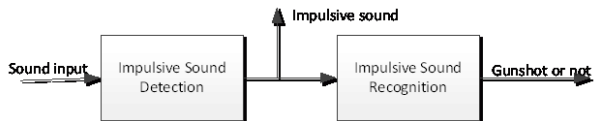

Figure-1 Impulsive Sound Detection and Gunshot Recognition System

After detection of an impulsive sound, the detected window is transmitted to gunshot recognition phase. The impulsive sound detection system is works all the time, gunshot recognition system works after detection.

In earlier works high detection and recognition rates were achieved in silent environment but in noisy environments these rates are low and false alarm rates are high [1]. In this paper noise effect is reduced, so in noisy environments high detection rates are achieved. This is mainly due to our new formulation of energy to find impulsive sounds.

The remainder part of the paper is organized as follows. In the second part literature review was explained, gunshot acoustics is explained in the third part, later in the fourth part our proposal for impulsive sound detection and recognition of gunshots were explained. At last section comes conclusion.

## II. LITERATURE REVIEW

### A. Commercial Systems

Currently there are several commercial gunshot detection and localization systems. But it is still an interesting topic in military area on which research is done.

The most widely known gunshot detection system is the Boomerang which is developed by Raytheon BBN Technologies. US army used it on Hummer vehicles in Iraq [2].

Shotspotter, can detect gunshots in a city and inform the police stations [3]. It is installed in many cities in US and they are used actively.

Pinpoint [4] and SWATS [5] (Shoulder Worn Acoustic Targeting System) systems can detect sniper shots and their location by way of the sensors installed on the shoulder of soldiers. There is not any paper or report about the signal processing algorithms and their performance.

### B. Detection and Recognition Algorithms in Literature

There are several techniques proposed for detection and recognition of impulsive sounds: In [6] for detection of impulsive sound average energy is calculated in a given T period. If energy is above a limit value it is decided that there is an impulsive sound in that period. Limit value is related with the noise in the environment and the coverage of the system. In [6], for gunshot recognition MFCC (Mel-Frequency Cepstral Coefficients) and LPC (Linear Predictive Coding) features are used. Cross-correlation and SVM (Support Vector Machine) are used for 99.7% correct recognition rate.

In [7] it is explained how to select a sub feature set which will give the best result for gunshot recognition, from a total 47 features. After selection of features two parallel working GMM (Gaussian Mixture Model) are used. In this study gunshot from noise and scream from noise are tried to be discriminated. In high noise environment 90% success was achieved.

In [8] MFCC and LPC features are extracted, HMM (Hidden Markov Model) and correlation against a template methods are used for classification. A database of sounds which consists of gunshots, balloon explosions, handclaps and speech sounds is used for classification. Although correlation against templates is a simple algorithm, it has given better results than the HMM.

In [9], [10], [11] and [12] other different methods for the recognition of gunshots are described.

### III. ACOUSTICS OF GUNSHOT

For the detection of gunshot, gunshot acoustics should be known in advance. In [13] it is given detailed explanation about the gunshot acoustics.

#### A. Muzzle Blast

A conventional firearm uses confined explosive charge to propel the bullet out of the gun barrel. The sound of the explosion is emitted from the gun in all directions, but the majority of the acoustic energy expelled in the direction of the barrel is pointing. The explosive shock wave and sound energy emanating from the barrel is referred as muzzle blast and typically lasts less than 3-7 msec. The muzzle blast acoustic wave propagates through the air at the speed of sound and interacts with the surrounding ground surface, obstacles, temperature and wind gradients in the air, spherical spreading, and atmospheric absorption. If a recording microphone is located close to the firearm, the direct sound of the muzzle blast is the primary acoustical signal. On the other hand, if the microphone is located at a greater distance from the firearm the direct sound path may be obscured and the received signal will exhibit propagation effects, multipath reflections and reverberation [13].

#### B. Supersonic Projectile

In addition to muzzle blast, another gunshot information is present if the bullet travels at supersonic speed. The supersonic projectile's passage through the air launches an acoustic shock wave propagating outward from the bullet's path. The shock wave expands in conic fashion behind bullet, with the wave front propagating outward at the speed of sound [13].

### IV. PROPOSED METHOD

#### A. Analysis of Current Methods

Mainly there are two types of methods which is used to recognize gunshots. First type of methods is based on feature extraction from sound signals. These features are used to train a machine learning algorithm by using a training set. By using a test set the success of machine learning algorithm and features are tested. There is no separate impulsive sound detection part. In these methods for the training and test set features are extracted in the same conditions. Although results show high success rates for those tests, if we extract features in different conditions then the success rates are low. Noise level, geographic location and climate can be examples of these conditions.

The second type of methods first detects impulsive sounds, and then found window is given to recognition part. Recognition part is the same as the first type of methods. In detection phase they generally use energy or power and try to eliminate noise by way of threshold level. These are prone to the same class of detection hazards as the first type of methods have, because impulsive sound comprises not just signature of the gunshot but the impulse response of the environment.

#### B. Detection of Impulsive Sound

As it is explained in gunshot acoustics, a gunshot lasts 3-7 msec. After this time we hear reverberations and echoes. For this reason we determined a window size of 6 msec to search for impulsive sound. We used in our experiments a sampling frequency of 16 KHz. The sample size of the window to search for an impulsive sound which is caused by a gunshot is Fs (sampling frequency in KHz) * 6/1000.

Proposed impulsive sound detection algorithm uses the expected value and the variance of the high frequency components of the sound in the window. The Fourier transform of the sound in the window by using proposed window sample count (this approximated to 99) is calculated. After taking the Discrete Fourier Transform 5-8 KHz (if the sound is sampled at 16 KHz) components of the signal are used to calculate the expected value and the variance. If the expected value is greater than 0.5 and the variance is greater than 0.2 we decide that this window consists impulsive sound. This calculation lasts at the end of buffer memory. We also make normalization thus the amplitude of the signal in the buffer is at most 1 or -1. After normalization of the buffer, we do the steps in Algorithm 1.

| Algorithm 1 Impulsive sound detection (Fs=16 KHz) |
|---|
| 1. Let N = 99, k = 0, l (window index) = 0 |
| 2. Calculate Discrete Fourier Transform of the window $X_l[k] = \sum_{n=0}^{N-1} x[n+l]e^{-j2\pi nk/N}$ k = 0,1,...,N-1 |
| 3. Calculate $E[|X_l[k]|]$ for $30 <= k <= 49$ |
| 4. Calculate $var[|X_l[k]|]$ for $30 <= k <= 49$ |
| 5. If ($E[|X_l[k]|] > 0.5$ and $var[|X_l[k]|] > 0.2$) then Give $X_l[k]$ for $30 <= k <= 49$ to recognition phase |
| 6. Let l = l +1, repeat from 2. |

Firstly we derived above formulas by making experiments. Actually in the above formulas we are calculating the energy of the sound signals but in a different way. In discrete time and frequency domain we calculate energy respectively as follows:

$$E = \sum_{n=0}^{N-1} |x[n]|^2 \quad (4)$$

$$E = 1/N \sum_{k=0}^{N-1} |X[k]|^2 \quad (5)$$

Variance of a sequence is calculated as follows:

$$Var(X[n]) = E[X^2[n]] - E[X[n]]E[X[n]] \quad (6)$$

$E[X^2[n]]$ component is expected value of energy. As seen, expected value of the energy is equal to addition of variance of the signal plus square of expected value of the signal. In (2) and (3) energy of sound signal is checked. We think that impulsive sound should contain variance so energy formula should also contain variance.

Impulsive sounds were detected after they were inserted into various noise. In Table-1 it is shown that these noise types, SNR (Signal to Noise Ratio) levels and the detection values of expected value and variance. SNR is calculated as proposed in [1]:

$$SNR = 10.\log_{10} \frac{\sum_{k=0}^{N-1} |x[k]|^2}{\sum_{n=0}^{N-1} |y[n]|^2} \quad (7)$$

| Noise Type | Expected Value | Variance | False Alarm | SNR (dB) |
|---|---|---|---|---|
| Crowd | 0.549 | 0.215 | 3 | 10.45 |
| Helicopter | 0.563 | 0.203 | 0 | -15.38 |
| Airport | 0.613 | 0.270 | 0 | -5.47 |
| Train | 0.608 | 0.200 | 0 | 11.77 |
| Thunder | 0.514 | 0.234 | 0 | 1.82 |

Table-1 Detection of impulsive sound in various noise types.

### C. Recognition of Gunshot

After detection of impulsive sound as explained before, our aim is to decide if this impulsive sound is a gunshot or not. We extracted two features from the window which contains impulsive sound. These are frequency information and MFCC values. To test recognition phase, these quantities are compared with the other gunshots and impulsive sounds from which we extracted same features before. Comparison was made by WEKA tool [14].

### D. Tests

By using 25 microphones, balloon explosions and pseudo gunshots are recorded in an open area. A pseudo gunshot recording is shown in Figure-2. There are no reflecting surfaces nearby except the earth floor. Thus the recording is near to ideal impulsive sound. These recordings embedded into a randomly selected point in various kinds of noises such as traffic, thunder, train, crowd etc. found in the Internet. Then using our method described before impulsive sounds are detected. In Figure-3 it is shown a pseudo gunshot recording embedded into crowd noise. The results of these tests are shown in Table-1. All the embedded impulsive sounds are found at high noise level. The SNR level at embedding point is also shown in the Table-1.

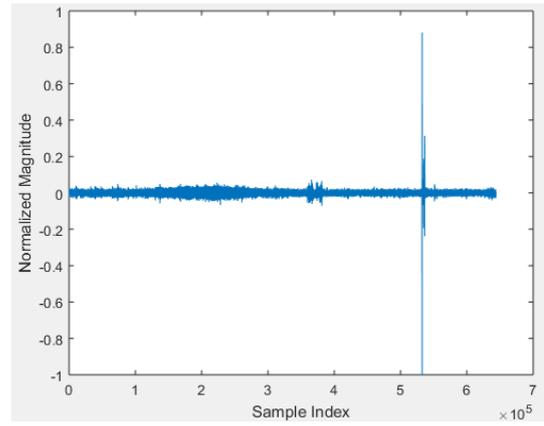

Figure-2 Pseudo gunshot sound

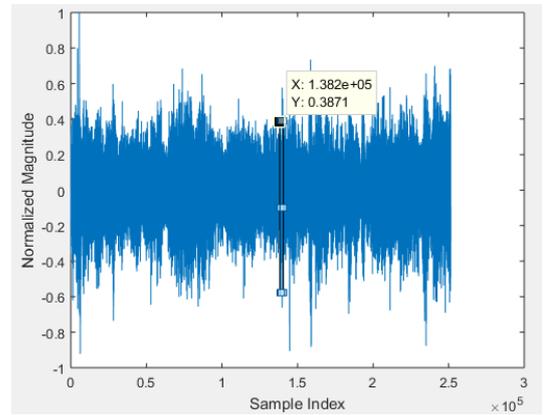

Figure-3 Pseudo gunshot sound embedded into crowd noise

| Feature | Algorithm | True Positive Rate | False Positive Rate |
|---|---|---|---|
| High Frequency Amplitude | DVM (libSVM) | 87.4 | 14 |
| MFCC | DVM (libSVM) | 94.7 | 5.6 |
| High Frequency Amplitude | MultiLayer Perceptron | 86.1 | 14.9 |
| MFCC | MultiLayer Perceptron | 92.6 | 7.7 |

Table-2 Gunshot recognition rates

To test the gunshot recognition phase of our system, we used the real gunshot recordings found at [15]. These gunshot recordings compared against or own recording of balloon and pseudo gunshot. The frequency and the MFCC values of the window which contains impulsive sound are used as features. To extract these features WEKA [14] tool is used. Later these features are fed to machine learning algorithms. Best result is obtained when DVM and MFCC are used together. Table-1 shows the results of these tests. 8 fold cross-validation method was used. When MFCC values were used, real gunshot recording could be discriminated from pseudo gunshot and balloon explosions with 95% success rate.

## V. CONCLUSION

In this paper we proposed a new method for detection of impulsive sounds. We based on theoretical analysis of gunshot acoustics and decided on a window size. After making experiments understood how to find impulsive sounds and described the reason. Mainly we present a new energy formula which is taking into account mean and variance of the signal sequence.

Our gunshot recognition phase is similar to previous works as a method. But mainly due to our window size, the sound samples and the frequencies we use to extract features, our recognition phase gives better results.